\documentclass[11pt]{amsart}

\usepackage{amsmath,amssymb,amsfonts,fullpage,amsbsy,graphicx}
\usepackage{equation,color}
\usepackage[dvipsnames]{xcolor}
\usepackage{amsaddr,enumerat,qtree}
\usepackage{ulem}
\usepackage{lineno}

\usepackage[authoryear]{natbib}

\newtheorem{theo}{Theorem}

\newtheorem{prop}[theo]{Proposition}

\newtheorem{defn}[theo]{Definition}
\newtheorem{rem}[theo]{Remark}

\newcommand{\R}{{\mathbb R}}

\newcommand{\bemat}[1]{\left( \begin{array}{#1}}
\newcommand{\emat}{\end{array} \right)}                                

\newcommand{\seq}[1]{\left(#1\right)_{m=1}^\infty}

\title{Decomposition of the Leinster-Cobbold Diversity Index} 

\author{Bingzhang Chen, Michael Grinfeld} \address{Department of
  Mathematics and Statistics, University of Strathclyde, 26 Richmond
  Street, Glasgow G1 1XH, UK}

\email{bingzhang.chen@strath.ac.uk,m.grinfeld@strath.ac.uk}

\newcommand{\mb}[1]{{\boldsymbol #1}}

\newcommand{\blot}[1]{}

\newcommand{\U}{\mathcal{U}}

\renewcommand{\S}{{\mathbb S}}
\newcommand{\B}{{\mathbb B}}
 
\begin{document}


\begin{abstract}
  The Leinster and Cobbold diversity index \citep{LeinsterCobbold2012}
  possesses a number of merits; in particular, it generalises many
  existing indices and defines an effective number. We present a
  scheme to quantify the contribution of richness, evenness, and
  taxonomic similarity to this index. Compared to the work of
  \citep{vanDam2019}, our approach gives unbiased estimates of both
  evenness and similarity in a non-homogeneous community. We also
  introduce a notion of taxonomic tree equilibration which should be
  of use in the description of community structure.

    \bigskip 

\end{abstract}

\maketitle

\section{Introduction}

Measuring biodiversity is a difficult task due to sampling issues and
accounting for missing data, but also as there is no one universally
accepted definition of what biodiversity is \citep{Daly2018}.  In
ecological practice, definitions of biodiversity can include
contributions from multiple channels of information such as the number
of species (``richness''), dominance or rarity relations among the
constituent species (``evenness''), and measures of ``similarity''
among the species (estimated either from taxonomic or phylogenetic
relationships, or from functional traits relationships)
\citep{Purvis2000, LeinsterCobbold2012}. Biogeographic patterns of
diversity can depend on the definitions used. For example,
\citep{Stuart2013} showed that biodiversity hotspots can shift from the
tropics to higher latitudes if one only considers abundances or also
takes account of functional traits similarity of species.

An outstanding challenge in the field of conservation ecology is to
relate the various aspects of biodiversity data to the functioning of
ecosystems \citep{Maureaud2019, Hillebrand2018}.

Thus the goal is to construct a biodiversity index that would carry
information about as many aspects of diversity as possible. This goal
has been actively pursued \citep{Rao1982, LeinsterCobbold2012,
  ChaoChiuJost2014}. By ``carrying information'' we means that, for
example, we should be able to extract information about richness or
evenness from our index. One way of extracting such information is to
decompose the index additively or multiplicatively into components
that can be interpreted in a biologically meaningful way; see for
example discussions of $\alpha$- $\beta$- and $\gamma$- diversity
\citep{Jost2007, Anderson2011}. A priori it is not clear why such a
decomposition would exist, whether it has to be unique, and in cases
of non-uniqueness, what are the conditions for optimality of a
decomposition.

As an example of this approach, van Dam \citep{vanDam2019} has recently
proposed a straightforward decomposition of the Leinster--Cobbold (LC)
index \citep{LeinsterCobbold2012}. In a sense, our work below is a
generalisation of the work of van Dam, which uses intrinsic properties
of the LC index to remove an important bias in van Dam's decomposition
with intriguing and far-reaching consequences.

The structure of the paper is as follows. In Section~\ref{Concepts} we
discuss the definitions of richness, evenness and similarity using
\citep{ChaoChiuJost2014,ChiuJostChao2014,Daly2018,GregoriusGillet2022}.
In Section~\ref{LC} we collect the required information about the LC
index following \citep{LeinsterCobbold2012, LeinsterMeckes2016}; it
subsumes many other diversity indices such Rao's index that is widely
used in functional ecology \citep{Rao1982, RicottaMoretti2011}. In
Section~\ref{Scheme} we present van Dam's and then our decomposition
and its consequences. Finally, in Section~\ref{Remarks} we discuss the
relation of our work to that or Chao and
Ricotta~\citep{ChaoRicotta2019}, extensions and open problems.

\section{Diversity components} \label{Concepts}

\subsection{Notation}

First of all, we need to establish notation. Everywhere below we
assume that the number of species in a community is fixed at $n>1$.

We will use $\mb{p}=(p_1,\, \ldots, p_n)$ to denote the vector of
relative abundances, and let 
\begin{equation}\label{simp}
  \Delta(n) = \{ \mb{x} \in \R^n,\; | \; x_k \geq 0, \; \; k= 1, \ldots,
  n, \; \; \sum_{k=1}^n x_k =1\}
\end{equation}
be the standard $n-1$-simplex in $\R^n$.

\begin{rem}
  It has to be emphasised that admissible relative abundance vectors
  $\mb{p}$ take values in $\Delta(n)^\circ$, the interior of
  $\Delta(n) $:
\begin{equation}\label{simpo}
  \Delta(n)^\circ = \{ \mb{x} \in \R^n,\; | \; x_k > 0, \; \; k= 1, \ldots,
  n, \; \; \sum_{k+1}^n x_k =1\},
\end{equation}
which is not a closed set in $\R^n$; the consequence of that is that
there are converging sequences $\seq{\mb{p}_m}$ in $\Delta(n)^\circ$,
whose limit is contained in the boundary $\partial \Delta(n) = \Delta(n)
\backslash \Delta(n)^\circ$; such limits by necessity correspond to
communities with fewer than $n$ species.
\end{rem}

We will often use the vector
\begin{equation}\label{vect}
\mb{p}_h = \left( \frac1n, \ldots, \, \frac1n \right) \in
\Delta(n)^\circ;
\end{equation}
the subscript $h$ stands for ``homogeneous''.  $\mb{p}_h$ is the
relative abundance of a community where each species is represented
equally. We denote by
$M(n) \subset \Delta(n) \backslash \Delta(n)^\circ$ the set of
$n$-vectors having one component equal to 1 and the rest equal to
zero. Thus, $\mb{m} \in M(n)$ is the relative abundance vector of a
monomorphic community. $\bf{1}$ will stand below for an $n$-vector
with all components equal to $1$.

Next, we need to discuss sets of $n \times n$ matrices. First of all,
we will denote the $n \times n$ identity matrix by $I_n$. We will use
the notation $J_n$ for the $n \times n$ matrix of ones.

In the present paper, for simplicity, we will be working with
ultrametric matrices; this choice is motivated by the fact that
similarity matrices (see subsection~\ref{dlc}) constructed using
taxonomic trees are necessarily ultrametric and since using them
simplifies the theory of \citep{LeinsterMeckes2016}. For more
information on ultrametric matrices, please see
\citep[Ch. 3]{Della2014} and \citep{Leinster2013,
  LeinsterMeckes2016}. We will denote that set of all ultrametric
$n \times n$ matrices by $\U(n)$ and its interior by $\U(n)^\circ$.

\begin{defn}[Defn. 3.2 of \citep{Della2014}] \label{ultra}
  A symmetric $n \times n$ matrix $A$ is {\bf ultrametric} if $A_{i,i}
  \geq \max_{k \neq i} A_{ik}$, $i, k \in \{1,\ldots, n\}$ and $A_{ik}
  \geq \min \{ A_{ij}, A_{jk} \}$, $i,j,k \in \{1,\ldots, n\}$.    
\end{defn}

\begin{rem}
  Note that in \citep[Example 12]{LeinsterMeckes2016} Leinster and
  Meckes take the matrices they call ultrametric to be strictly
  diagonally dominant. That would preclude the set of ultrametric
  matrices from being closed; so in our definition $J_n \in \U(n)$.
\end{rem}

\subsection{Evenness}

The concept of evenness (for which see,
e.g. \citep{ChaoChiuJost2014,ChaoRicotta2019,GregoriusGillet2022}) and
references therein), is rather problematic. First of all, the
terminology is badly chosen as it would immediately seem that the
``most even'' population of $n$ species is one for which the vector of
relative abundances is the homogeneous vector $\mb{p}_h$, i.e. one
where every species is equally represented.  Thus, like the case of
richness discussed below, the terminology seems to be precluding
discussion. As rightly pointed in \citep{GregoriusGillet2022}, such a
categorical answer to the question of maximal evenness leaves open the
discussion of what would constitute ``maximum unevenness'' in
$\Delta(n)^\circ$.  van Dam \citep{vanDam2019} uses instead the concept
of ``balance'', which seems to us a better term; this is the concept
which, after defining it properly (see (\ref{bal})) we will be using
below.

\subsection{Richness}

Richness is sometimes summarily defined to be the number of species,
see e.g. \citep{Daly2018}. Our approach below allows us to retain this
definition but at a price. Such a definition is open to the same
criticism as the notion of maximal evenness defined by $\mb{p}_h$ that
we have discussed above. It is again ``species-centric'', and takes
into account only the last level of taxonomic classification. Below,
in Section~\ref{Scheme} we suggest how to introduce a defensible new
notion of richness that uses taxonomic information.

\subsection{Similarity}\label{sim}

In this section we discuss the construction of taxonomic similarity
matrices $Z$ for a community with $n$ species. 

\begin{figure}[h]
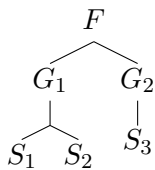

\hspace*{-2in} \Tree [.$F$ [ .$G_1$ [ $S_1$ $S_2$ ] ] [.$G_2$ $S_3$ ] ] 
\caption{A taxonomic tree}\label{tree1}
\end{figure}

The usual way of constructing similarity matrices $Z$ which are
automatically ultrametric is to assign distances between different
levels of a taxonomic tree. Then the taxonomic distance $d(i,j)$
between two species is the sum of distance from the nodes
corresponding to these species to the first common node, and then one
puts $Z_{ij}=e^{-d(i,j)}$ or if the maximal distance in the tree has
been normalised to 1, one could put $Z_{ij}=1-d(i,j)$. As a example,
consider the tree in Figure~\ref{tree1}

and set the species-genus and the genus-family distance to be
$0.3$. If we use the additive recipe, we get
\begin{equation}\label{zex}
  Z_1 = \bemat{ccc} 1 & 0.7 & 0.4\\
                  0.7 & 1   & 0.4\\
                  0.4 & 0.4 &  1
       \emat           
\end{equation}

\section{The Leinster--Cobbold diversity index}\label{LC}

In their influential paper \citep{LeinsterCobbold2012}, Leinster and
Cobbold introduced a far-reaching generalisation of Hill numbers, for
discussions of which see \citep{ChiuJostChao2014,Hill1973}, the LC
index. For details on its properties, see
\citep{LeinsterCobbold2012,LeinsterMeckes2016}; here we just collect
the bare minimum in the framework of taxonomic (ultrametric)
similarity matrices.

\subsection{Definition of the LC index} \label{dlc}

As in the definition of Hill numbers, below $q \in [0,\infty)$ is the
sensitivity parameter, measuring the importance given to rare
species. Then for a community of $n$ species with relative
abundance vector $\mb{p}$ and (ultrametric) similarity matrix $Z$, we
have
\begin{defn}\label{LCi}
  The {\bf LC diversity of
  order $q$} is
\begin{equation}\label{lcdef}
  F(Z,\mb{p},q) :=\left(\sum_{i=1}^n p_i \left(Z
      \mb{p}^T\right)_i^{q-1}\right)^{1/(1-q)}. 
\end{equation}
\end{defn}

Note that \citep{LeinsterCobbold2012} use a different notation, similar
to the Hill number notation in the literature; they denote the
right-hand side of (\ref{lcdef}) by ${}^qD^Z(\mb{p})$. We prefer the
notation used here as it clearly shows functional dependencies and
allows easy generalisation, which we discuss briefly in
Section~\ref{Remarks}. We collect the required properties of the LC
index in the proposition below and in subsection~\ref{max}.

\begin{prop} \label{LCprop}
  Let $\mb{p}\in \Delta(n)^\circ$. $Z \in U(n)$. Then 
  \begin{enumerate}[(a)]
  \item $F(\mb{p},Z,q)$ is a monotone decreasing function of $q$;
  \item $F(\mb{p},Z,q) < F(\mb{p}, I_n,q)$ for all $q$ if $Z \neq I_n$;  
  \item $F(\mb{p}_h,I_n,q)=n$ for all $q$;
  \item $F(\mb{p},J_n,q)=1$ for all $q$.
  \end{enumerate} 
  \end{prop}   

 For proofs of (a) and (b) please see \citep{LeinsterCobbold2012}; the rest
 are immediate.

 Following \citep{LeinsterMeckes2016}, we now discuss the concept of a
 \emph{maximally balanced abundance vector} for a community of $n$
 species with an ultrametric similarity matrix $Z$. 
 
\subsection{A crucial property  of the LC index}\label{max}

Using only ultrametric taxonomic similarity matrices simplifies the
presentation considerably. For the more general case where the
similarity matrix is simply a symmetric matrix with positive elements,
see~\citep{LeinsterMeckes2016}. The results of that paper have not, in
our opinion, been sufficiently seriously considered by the
biodiversity community.

We present two theorems from \citep{LeinsterMeckes2016}.  First of all
we have the following existence and uniqueness result for maximisers
of the LC diversity index.

\begin{theo}\label{exmax}
  For each $Z \in \U(n)^\circ$ there exists a unique abundance vector
  $\mb{p}^* \in \Delta(n)^\circ$ that maximises $F(Z,\mb{p},q)$ for
    every value of $q \in [0, \infty)$.
\end{theo}

\begin{defn}\label{mb}
Given $Z \in \U(n)^\circ$, we call the corresponding abundance vector
$\mb{p}^*$ the {\bf maximally balanced} abundance vector. 
\end{defn}

This is the vector that corresponds to $\mb{p}_h$ that arises in
theories that do not take into account taxonomic similarity.

Computing the maximally balanced vector in the case of ultrametric
similarity matrices is a simple matter of solving a system of linear
equations and normalising. If the similarity matrix is not
ultrametric, the situation is more complex; see
\citep{LeinsterMeckes2016} for details.

\begin{theo}\label{comp}
  Given $Z \in \U(n)^\circ$, $\mb{p}^*$ is given by
  \[
    p^*_i = \frac{w_i}{\sum_{j=1}^n w_j},
  \]
where $\mb{w}$ solves the system of equations $Z\mb{w} =
\mb{1}$, where $\mb{1}$ is a column vector of ones.    
\end{theo}

Note that \citep[Lemma 6]{LeinsterMeckes2016} provides an alternative
way of computing $\mb{p}^*$.

\begin{defn}\label{pt}
  A taxonomic tree will be called {\bf taxonomically equilibrated} if
  $\mb{p}^* = \mb{p}_h$.
\end{defn}

Of course we have 

\begin{prop}\label{balance}
  If at each level of the tree all the nodes have the same degree,
  the taxonomic tree is equilibrated.
\end{prop}

The converse of Proposition~\ref{balance} does not hold, i. e. there
are taxonomic graphs that do not satisfy the conditions of that
proposition, for which a metric $d(\cdot,\cdot)$ can be assigned such
that the resulting $\mb{p}^*$ is $\mb{p}_h$. An example is provided by
the following tree:

  \begin{figure}[h]
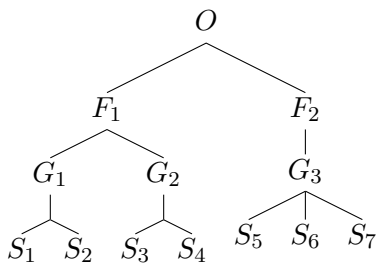

  \hspace*{-2in} \Tree [ .$O$ [ .$F_1$ [ .$G_1$ [ $S_1$ $S_2$ ] ]
  [ .$G_2$ [ $S_3$ $S_4$ ] ] ]
  [ .$F_2$  [.$G_3$ $S_5$ $S_6$ $S_7$ ].$G_3$ ] ]   
  \caption{A taxonomic tree that allows $\mb{p}^*=\mb{p}_h$.}
  \label{tree2}
\end{figure}

\begin{figure}[h]
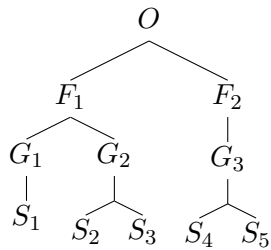

  \hspace*{-2in} \Tree [.$O$ [ .$F_1$ [ .$G_1$ $S_1$ ]
  [.$G_2$ [$S_2$ $S_3$ ] ] ] [.$F_2$ [.$G_3$ [$S_4$ $S_5$ ] ] ] ] 
  \caption{A taxonomic tree for which $\mb{p}^* \neq \mb{p}_h$ is
    guaranteed.}\label{tree3}
\end{figure}

It is not hard to show that the assignment of species-genus,
genus-family and family-order distances of $0.25$ and using the
additive recipe, results in a similarity matrix for which
$\mb{p}^*=\mb{p}_h$. Thus there is a trichotomy of taxonomic trees:
those in Proposition~\ref{balance} is which $\mb{p}^*=\mb{p}_h$ holds
for every assignment of distances; those where such assignments can be
chosen, as in Figure~\ref{tree2}, and such that no assignment of
distances results in a homogeneous maximally balanced abundance
vector; an example of such a tree is in Figure~\ref{tree3}.

We will discuss this trichotomy in more detail in \citep{ChenGrinfeld2023}.

\begin{rem} Compared to the diversity metrics proposed by Chao {\em
  et al.}  \citep{ChaoChiuJost2014}, the LC index has the flexibility
to take into account taxonomic, phylogenetic, and functional diversity
simultaneously. However in that case the resulting similarity matrices
are no longer ultrametric and we leave that more general case for
future study. Also note that the above trichotomy of taxonomic trees
would not necessarily exit if there were a canonical way of constructing
similarity matrices. 
\end{rem}


\section{An unbiased decomposition scheme}
\label{Scheme}

We are now ready to propose a decomposition scheme for the LC
index. It is best to start with the decomposition scheme proposed by
van Dam \citep{vanDam2019} and see why it has to be modified. van Dam
writes
\begin{equation}\label{dec1}
  F(\mb{p},Z,q) = \frac{F(\mb{p},Z,q)}{F(\mb{p},I_n,q)} \cdot
  \frac{F(\mb{p},I_n,q)}{F(\mb{p}_h,I_n, q)} \cdot F(\mb{p}_h,I_n, q).
\end{equation}
The first fraction is clearly a measure of dissimilarity, while the
second fraction is a measure of balance; of course
$F(\mb{p}_h,I_n, q)=n$, so the last term in the right-hand side is a
richness. From many points of view, this is a good decomposition as
the two fractions always lie in the interval $(1/n,1]$. Consult
Proposition~\ref{LCprop} to see that the infimum $1/n$ is never
reached. The problem here is with the definition of the measure of
balance, as it does not take into account the taxonomic similarity
matrix $Z$ while the dissimilarity measure uses information from both
$\mb{p}$ and $Z$. We call such a decomposition {\bf asymmetrically
  biased}. A different asymmetrically biased  decomposition is
given by
\begin{equation}\label{dec2}
  F(\mb{p},Z,q) = \frac{F(\mb{p},Z,q)}{F(\mb{p}_h,Z,q)} \cdot
  \frac{F(\mb{p}_h,Z,q)}{F(\mb{p}_h,I_n, q)} \cdot F(\mb{p}_h,I_n, q).
\end{equation}
In this decomposition the first fraction is a measure of balance, the
second a measure of dissimilarity, while as before, the last term is
richness. Of course here it is the measure of dissimilarity that is
asymmetrically biased. A possibility that might be considered of
multiplying (\ref{dec1}) and (\ref{dec2}) and taking the square
root. That will give us a decomposition which we would call {\bf
  unbiased} (though it could be also called ``symmetrically biased'')
. However, there is an additional problem in (\ref{dec2}) which is
that the first fraction in the right-hand side can take values larger
than one if for example, $Z$ is not a similarity matrix of a
taxonomically equilibrated tree and $\mb{p}=\mb{p}^*$. We do not
pursue this direction as we do not see any reason for a relative
measure not to take values in $[0,1]$.

The price of ensuring normalisation is having to deal with richness in
more detail. Consider instead of (\ref{dec2}), the following decomposition:
\begin{equation}\label{dec3}
  F(\mb{p},Z,q) =\frac{F(\mb{p},Z,q)}{F(\mb{p}^*,Z,q)}
    \cdot
  \frac{F(\mb{p}^*,Z,q)}{F(\mb{p}^*,I_n, q)}
     \cdot F(\mb{p}^*,I_n, q).
\end{equation}
It is asymmetrically biased as second factor does not involve
$\mb{p}$. We will discuss the interpretation of $F(\mb{p}^*,I_n, q)$ later.

To obtain an {\bf unbiased} decomposition, we therefore multiply
(\ref{dec1}) and (\ref{dec3}) and take a square root. The result is
\begin{equation}\label{dec4}
  F(\mb{p},Z,q) =\sqrt{\frac{F(\mb{p},Z,q)}{F(\mb{p}^*,Z,q)}
    \frac{F(\mb{p},I_n,q)}{F(\mb{p}_h,I_n, q)}} \cdot
  \sqrt{\frac{F(\mb{p},Z,q)}{F(\mb{p},I_n,q)}
    \frac{F(\mb{p}^*,Z,q)}{F(\mb{p}^*,I_n, q)}} \cdot
  \sqrt{n F(\mb{p}^*,I_n, q)}.
\end{equation}
The last term in the right-hand side can be rewritten as
\begin{equation}\label{equil} 
  \sqrt{n F(\mb{p}^*,I_n, q)}= \sqrt{ \frac{F(\mb{p}^*,I_n,q)}{n}} n
  := E(Z,q)n.
\end{equation}
The term $E(z,q)$ expresses the lack of equilibration in the taxonomic
tree; see Definition~\ref{pt} and Theorem~\ref{balance}.

Putting 
\begin{equation}\label{bal}
B(\mb{p},Z,q)  = \sqrt{\frac{F(\mb{p},Z,q)}{F(\mb{p}^*,Z,q)}
  \frac{F(\mb{p},I_n,q)}{F(\mb{p}_h,I_n, q)}},
\end{equation}
\begin{equation}\label{dis}
D(\mb{p},Z,q)  =   \sqrt{\frac{F(\mb{p},Z,q)}{F(\mb{p},I_n,q)}
  \frac{F(\mb{p}^*,Z,q)}{F(\mb{p}^*,I_n, q)}},
\end{equation}

We finally write our decomposition as
\begin{equation}\label{decf}
F(\mb{p},Z,q) = B(\mb{p},Z,q) D(\mb{p},Z,q) E(Z,q)n., 
\end{equation}
i.e. a product of measures of balance $B(\mb{p},Z,q)$, dissimilarity
$D(\mb{p},Z,q)$, (lack of) equilibration $E(Z,q)$ and the classical
richness $n$.

Note that by construction both the measure of balance (\ref{bal}) and
dissimilarity (\ref{dis}) are constrained to lie in $[0,1]$ by
Proposition~\ref{LCprop}.

Both the measure of balance and of dissimilarity are geometric means
of an unbiased measure and a biased one. It does not seem possible to
find a truly unbiased decomposition of the LC index, which is the
reason we could call the decomposition (\ref{decf}) symmetrically biased.

Note that though $B(\mb{p},Z,q) \geq 1/(\sqrt{F(\mb{p}^*,Z,q)}$ (the
right-hand side being independent of $q$), it is not clear what
vector $\mb{p}(q)$ maximises it for a particular value of $q$. Of
course the value $1$ is reached for $q=0$ by the choice
$\mb{p}(0)= \mb{p}^*$.

$E(Z,q)$ depends on $Z$, as $Z$ defines $\mb{p}^*$, and hence $E(Z,q)$
reflects the structure of the underlying taxonomic tree.  Note that in
the case of similarity matrices of taxonomically equilibrated trees,
for which we have $\mb{p}^*= \mb{p}_h$ and hence
$F(\mb{p}^*,I_n, q)=n$, so that $E(Z,q)=1$, If $Z$ does not correspond
to a taxonomically equilibrated taxonomic tree, $F(\mb{p}^*, I_n, q)$
is dependent on $q$.

\begin{rem}
  We could have defined a notion of ``richness'' by $R(z,q):=E(Z,q)n$,
  but the decomposition (\ref{decf}) seems to us more insightful and
  does not necessitate an advocacy of a new notion of
  richness. Definition~\ref{pt} singles out a class of taxonomic trees
  for which the two notions of richness coincide.
\end{rem}

\section{Practical examples} \label{Examples}
 
To illustrate how our decomposition approach differs from the ``ABC"
approach suggested by van Dam \citep{vanDam2019}, we use a simple
example in Leinster and Cobbold \citep{LeinsterCobbold2012} (their
Example 3; original data from \citep{deVries1997}). This example has
the nice feature that the difference between the two communities
changes the sign when $q$ increases from 0 to 2.
 
\begin{figure}
   \centerline{\includegraphics[width=12cm]{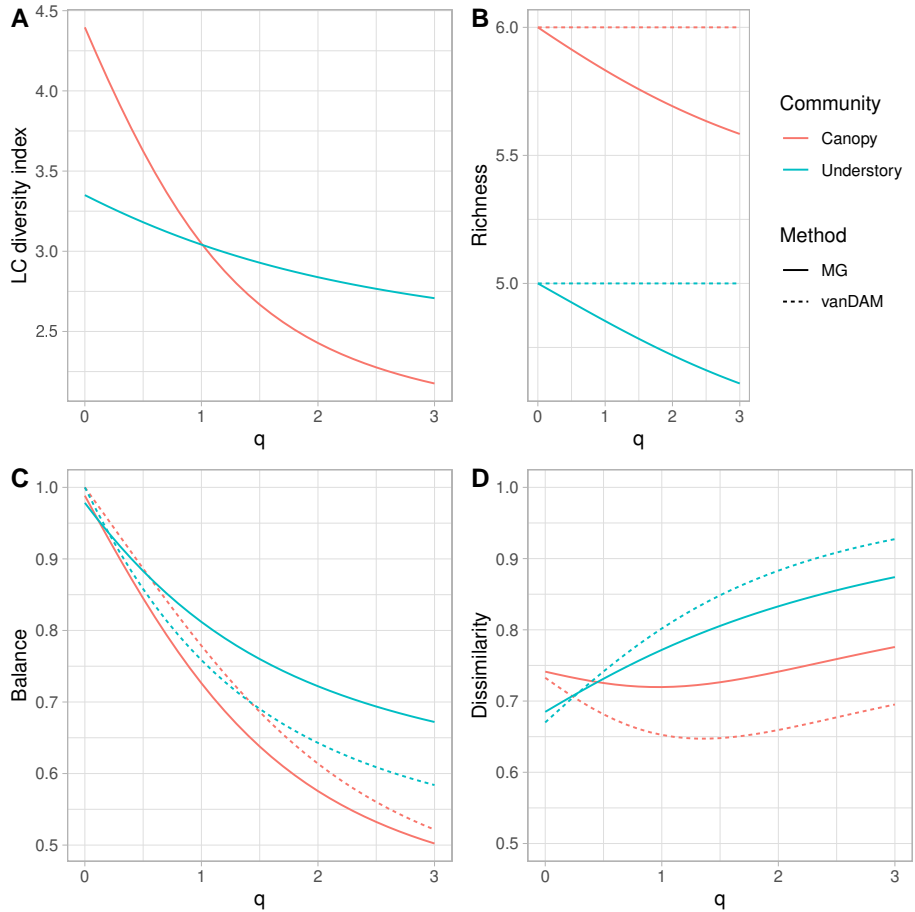}}
   \caption{Comparing our approach (``This study") and van Dam
     \citep{vanDam2019} to decompose (A) the LC diversity index into
     (B) Lack of equilibration, (C) Balance (or Evenness), and (D)
     Dissimilarity for the example of \textit{Charaxinae} in Leinster
     and Cobbold \citep{LeinsterCobbold2012}.}
  \label{fig1}
\end{figure}

Compared to van Dam \citep{vanDam2019}, our new decomposition approach
yields different interpretations regarding what aspects of diversity
lead to the differences in diversity estimates between the two
communities (Canopy vs. Understory). When we put more emphasis on rare
species ($q < 1$), the Canopy community is more diverse because it has
a larger number of species richness (6 vs. 5) and its species are
slightly more dissimilar with each other (Fig. \ref{fig1}A,D). By
contrast, when we focus more on abundant species ($q > 1$), the
Understory community becomes more diverse because of its greater
balance (evenness) of dominant species (Fig. \ref{fig1}A,C). The
difference between our decomposition approach and van Dam
\citep{vanDam2019} is that our approach predicts a much larger
difference in Balance than van Dam \citep{vanDam2019} (and a smaller
difference in species dissimilarity) when we increasingly focus on
dominant species (larger $q$). Our approach also shows that as $q$
increases, the Understory community shows a stronger lack of
equilibration (i.e., deviation of $\mb{p}$ from $\mb{p}^*$) than the
Canopy community, albeit this difference is small.  In summary, we
would interpret that the greater diversity of the Understory community
when we focus on dominant species is because its dominant species are
more balanced than those in the Canopy community. On the contrary, the
interpretation would be that the dominant species are more dissimilar
to each other in the Understory community if we use the approach of
van Dam \citep{vanDam2019}.

\section{Discussion}
\label{Remarks} 

The LC index uses three ``information streams'': the number of species
(richness) $n$, the relative abundance vector $\mb{p}$ and the
similarity matrix $Z$. We could in theory consider a diversity index
$F(c_1, \; \ldots, c_m; \, q)$, where $c_1, \ldots, \; c_m$ are
information streams, sources of information about the structure of the
community, expressed as some mathematical objects (vectors, matrices
higher order tensors). We could then follow the decomposition process
of Section~\ref{Scheme}: find $m!$ biased decompositions, multiply
them together and take the $m!$-root. However, this is already
unwieldy in the case of $m=3$. But note that this procedure is
unnecessary as the LC theory has a lot of built-in flexibility in the
definition of similarity matrices. As explained in
subsection~\ref{sim}, one can define a similarity matrix by setting
$Z_{ij}=e^{-d(i,j)}$, where $d(i,j)$ is some suitably defined distance
between species $i$ and $j$. Hence incorporating more information
streams can be thought about as changing the distance function
$d(\cdot,\cdot)$; in the process of incorporating such information,
such as functional similarity, the ultrametricty of the similarity
matrix is lost; it is possible that the resulting function
$d(\cdot,\cdot)$ will no longer be a metric, becoming more generally a
divergence measure. The point is $\mb{p}$ and (a suitably redefined)
$Z$ dependence of a diversity index is sufficient to incorporate all
relevant information.

In \citep{ChaoRicotta2019}, Chao and Ricotta show how to quantify
evenness using divergence measures. It is useful therefore to consider
the LC diversity index (\ref{LCi}) in this context. As now we deal
with balance as opposed to evenness, we will denote the resulting
balance index by $B$. First of all, let us note that (\ref{bal}) cannot
give rise to an divergence measure-based index of balance as there is
no well defined upper bound for it for all $q$.  It is still of course
useful in providing an estimate of the balance contribution to
the LC diversity index. These two statements are not in contradiction.

Clearly, the LC index itself provides a divergence measure-based
estimate (index) of balance via
\begin{equation}\label{e1}
\B = \frac{F(\mb{p},Z,q)-1}{F(\mb{p}^*,Z,q)-1},
\end{equation}
where one could alternatively write $1=F(\mb{m},Z,q)$ where $\mb{m}$
is any vector in $M(n)$.

Concerning similarity indices, it again does not seem to be possible
to utilise $D(\mb{p},Z,q)$ of (\ref{dis}) to this end. On the other
had, the LC index itself provides a divergence measure based
similarity index by
\begin{equation}\label{d1}
\S=\frac{F(\mb{p},I_n,q)-F(\mb{p},Z,q)}{F(\mb{p},I_n,q)-F(\mb{p},J_n,q)}.  
\end{equation}
Again, here the value 1 is never reached over $\U(n)^\circ$. It is not
hard to show the following proposition: 

\begin{prop}
  The indices $\B$, $\S$ satisfy all the
  requirements in \citep{ChaoRicotta2019}.
\end{prop}

To conclude, we have proposed a novel decomposition of the LC index.
Compared to a previous version of decomposition due to van Dam
\citep{vanDam2019}, our approach estimates the balance and
dissimilarity of the community more comprehensively (e.g., we not only
estimate dissimilarity for a homogeneous community but also consider
the present vector of relative abundance). As such, we believe that
our inference is more robust when comparing balance and dissimilarity
among communities. In addition, we had by necessity to introduce a
notion of taxonomic tree equilibration (which turns out to be an
important concept in our on-going work on quantifying un-evenness
\citep{ChenGrinfeld2023}), which is another descriptor of a
biological community. We advocate the use of our decomposition
(\ref{decf}) as a ``maximally unbiased'' estimate of contributions of
balance and (dis)similarity to diversity.

\bibliographystyle{apa}
\bibliography{Diversity.bib}

\end{document}